# TOWARDS SUPPORTING SIMULTANEOUS USE OF PROCESS-BASED QUALITY APPROACHES

**Zádor Dániel Kelemen[1], Katalin Balla[1], Jos Trienekens[2], Rob Kusters[2]**

[1] Department of Control Engineering and Information Technology,
Budapest University of Technology,
SQI – Hungarian Software Quality Consulting Institute Ltd.
kelemen.daniel@sqi.hu, balla.katalin@sqi.hu

[2] Department of Technology Management,
Eindhoven University of Technology
J.J.M.Trienekens@tue.nl, R.J.Kusters@tue.nl

**Abstract:** In this paper we describe the first steps of a PhD programme, having the goal to develop a common meta-model for different software quality approaches and methods. We focus on presenting the structure of quality approaches emphasizing the similarities amongst them. Understanding the structure of quality approaches helps supporting organizations in using multiple quality approaches and methods in the same time.

## 1 Introduction

Several quality methods, standards and models have been developed in the last few decades to guide software developing organizations in defining and institutionalizing their processes. These approaches are essential in improving the company's own quality system, but each of them uses an own view on quality.

However, software companies (want to / are forced to) use more quality approaches simultaneously, they often struggle with interpreting them, due to different terminology and their point of view on quality.

In the day-by-day consultancy work, we experienced that software companies often implement quality approaches separately, without unifying or harmonizing the common elements. Problems connected to process interpretation and implementation usually come when companies have separated process descriptions for different quality approaches and methods. In this case, project managers have to choose between approaches. Due to the different standards, projects are focusing in different ways on quality. Some (eg. ISO 9001:2000) projects are focusing on measurement of customer satisfaction and customer relationship management but not on technical solution and product integration. Others (eg. in CMMI-based projects) may concentrate on requirements development, management and traceability, but not on handling the customer's property.

The situation may become more complicated, when the processes built on different quality approaches include different descriptions of same areas (eg. change management or measurement).

Our work has the scope to give a solution to the problem described. The meta-model would make use of elements found to be common in more quality approaches. In the first phase of our research we focused on studying modelling techniques and on understanding basic elements of some quality approaches, in order to be able to choose the common elements that would form the basis of the meta-model. Here we present results of our investigation.

Chapter 2 of this paper summarises the process modelling evolution, based on G. Cugola's and C. Ghezzi's point of view. Afterwards (chapter 3) we describe the base structure of the most used (software) quality approaches (as ISO 9001:2000, ISO 9004:2000, ISO 90003:2004, CMMI-DEV v1.2, ISO-IEC 12207-95 and ISO-IEC 15939-2002). We show a comparison of the elements of quality approaches mentioned with the elements of a process. Finally, in chapter 4 we present an idea for harmonising common areas of quality approaches. We conclude by presenting the results obtained in our investigation.

Project has been supported by BME (IT)[2] (BME Innovation and Knowledge Centre of Information Technology) within the frame of Peter Pázmány Programme, National Office for Research and Technology, Hungary.

## 2 Process Modeling

In "Software Processes: a Retrospective and a Path to the Future" [Cugola et al. 1998] G. Cugola and C. Ghezzi have shown (table 1.) the main steps of software process evolution starting from the early 60's. In table 1 strengths and weaknesses of lifecycle models (1), methodologies (2), formal development (3), automation (4), management and improvement (5) are shown.

After these approaches, a new era came for processes: process modelling and process programming. In process modelling there are several research works, like PML[1], Little JIL [Osterweil 1987] [Osterweil 2007], Oz[2] Endeavors, BPM or enterprise modelling [Wortmann et al. 2007].

---

[1] Process Modeling Languages – research area, introduced by Osterweil [Osterweil 1987]

[2] Oz and Oz Web – the first "decentralized" PSEE was developed at Columbia University.

Process modelling can be classified in several ways, eg. by architectures and modelling approaches.

The minimalist process modelling approach describes only the most important elements of processes, and it is easily understandable for people. The maximalist approach describes and validates the whole process model. Processes built in maximalist way can be processed by computers, but are harder to understand by humans.

From the architectural point of view top-down process approaches start from the idea to the implementation, bottom-up approaches try to model the manifested processes.

Table 1. – Evolution of software processes

| Eg. | Strength | Weakness |
|---|---|---|
| Waterfall model | Well structured, clear documentation | Idealised processes |
| JSD[3], JSP | Based on experiences from previous projects | Informal notation, increased paperwork |
| Program development by stepwise refinement | Transforms specification to correct implementation | Not scalable, applicable only for small programs |
| SDE[4]s | Automation of some areas of software production | Requirements specification, design decisions cannot be automated |
| ISO9001: 2000, CMMI, TSP, PSP | Indirect assurance of quality products | Increased bureaucracy |

We have chosen the following goal in the Ph.D work: to understand the structure of the process based quality approaches, and create a common meta-model in a minimalist way, which will be easily understandable for quality managers and project managers. Using this meta-model, processes could be built in a top-down or a bottom-up way.

---

[3] JSD - Jackson System Development

[4] SDE - Software Development Environment

## 3 Structure of Quality Approaches

Process-based quality approaches are often textually described, and in order to model them we need to know what their basic elements are.

In Hungary, the most used and "mandatory" quality approach is ISO 9001:2000[5] (Full title: "ISO 9001:2000 Quality management systems – requirements"). Besides ISO 9001:2000, most used approaches are the Capability Maturity Model Integration (CMMI) and (Automotive) SPICE (Software Process Improvement and Capability dEtermination, also known as ISO/IEC 15504). While software companies use CMMI, suppliers of multinational car factories prefer SPICE as a second approach.

Further well-known standards connected to software processes are ISO 9004:2000, ISO 90003:2004, ISO-IEC 12207-95 and ISO-IEC 15939-2002.

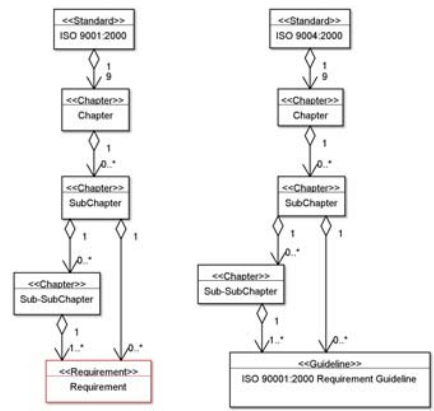

Fig. 1 - The Structure of ISO 9001:2000, ISO 9004:2000

ISO 9001:2000 is an international standard which contains general requirements for quality management systems (QMS). The requirements included in this standard are so general that can be applied at any company.

If we look at this standard, we can see that it contains 9 chapters, which could contain subchapters and the subchapters also can contain further subchapters. Requirements of this quality approach can be found at subchapter and sub-subchapter level in sentences. Figure 1. shows the structure of ISO 9001:2000 and two, other ISO 9001:2000-connected standards. The structure of ISO 9004:2000 "Quality management systems – Guidelines for performance improvements" and ISO/IEC 90003:2000 "Software Engineering – Guidelines for the application of ISO9001:2000 to computer software" are identical to ISO 9001:2000 because these are using ISO 9001:2000 as a basis, containing the same chapters. The only difference amongst them is that the latter two define guidelines instead of focusing on requirements.

As we already mentioned, other two widespread approaches are CMMI and SPICE. Here we focus on CMMI[6].

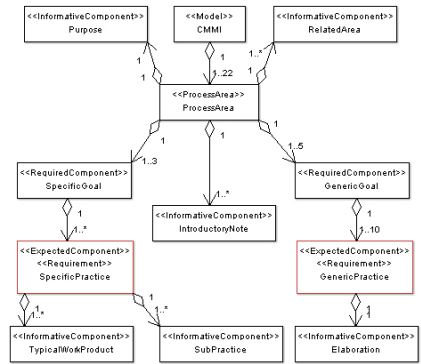

Fig. 2. - The Structure of CMMI-DEV v1.2

The actual version of CMMI, v1.2 defines 3 constellations: CMMI for Development, CMMI for Acquisition and CMMI for Services. Different constellations include different sets of process areas of the model. Looking at its structure, the model contains required, expected and informative components. Informative components are guidelines, specific and generic practices are the concrete, expected requirements. Required components are derived from expected components.

---

[5] See http://www.imcc.hu for the list of ISO 9001:2000 certified Hungarian (software) companies.

[6] CMMI is an integrated model, it integrates ideas from CMM, SPICE and further international quality standards, therefore most of the requirements of the SPICE model can be derived from CMMI.

Standard ISO/IEC 15939-2002 – "Information technology - Software measurement process" defines process activities and sub activities required for the measurement process.

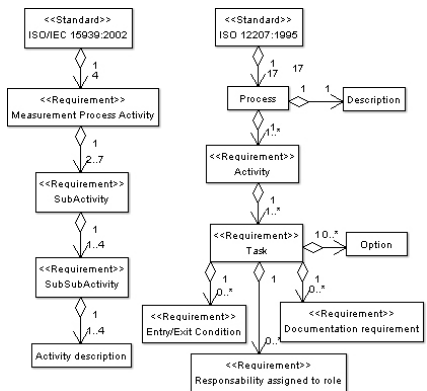

Fig. 3. - The Structure of ISO-IEC 15939-2002 and ISO-IEC 12207-95

ISO/IEC 12207-95 – "Information technology - Software life cycle process" describes processes, activities, tasks, entry and exit conditions, responsibilities and documentation requirements for software lifecycle processes.

## 4. Conclusion

In chapter 3. we have shown the structure of 6 different quality approaches. In each of them we found *requirements* or *guidelines*. In ISO standards, requirements and guidelines are usually textually described, in sentences. In other approaches, like CMM, CMMI or SPICE different levels and categories of requirements can be found. We started to analyse further approaches and methods like Agile methods and IT Infrastructure Library (ITIL) and we made similar observations.

We found in these approaches several types of elements: eg. chapter, requirements, guidelines, process, process description, activity, process activity, activity description, task, option, entry and exit condition, documentation requirement, responsibility, process area, specific and generic goal, specific and generic practice, typical work product, subpractice, practice elaboration etc.

Basic Process elements are *inputs*, *activities*, *outputs*, purpose, entry and exit criteria, roles, measures, and verification steps [SEI 2006].

Looking at the elements, we found several coincidences amongst elements of quality approaches and process elements. Process, process description, activity, process activity, activity description and task are proportional to the *activity* element of processes. Documentation requirements and typical work products are proportional to *inputs* and *outputs*.

We found element types which have no similarities to process elements. Such elements are eg. benefits, critical success factors, features or key performance indicators in ITIL.

Analysing the content of quality approaches we found that several approaches are focusing on the same problems (eg. change management can be found CMMI, ISO 9001:2000 and ITIL), but from different quality point of view.

In conclusion we can state that the idea to build a common meta-model for making the harmonisation of different of quality approaches and methods easier, seems both useful (as we emphasized in chapter 1) and feasible. The way towards such a meta-model starts by analysing modelling possibilities (chapter 2), and continues by identifying the elements of such a meta-model, starting from the analysis of quality approaches' structure (chapter 3).